\documentclass[twocolumn]{aastex63}
\turnoffedit

\newcommand{\sao}{\affiliation{Smithsonian Astrophysical Observatory, Cambridge, MA, USA}}
\newcommand{\umich}{\affiliation{University of Michigan, Ann Arbor, MI, USA}}

\newcommand{\csdl}{\affiliation{Charles Stark Draper Laboratory, Cambridge, MA, USA}}
\newcommand{\ucb}{\affiliation{University of California, Berkeley, CA, USA}}
\newcommand{\gsfc}{\affiliation{NASA, Goddard Space Flight Center, Greenbelt, MD, USA}}
\newcommand{\usra}{\affiliation{Universities Space Research Association, Huntsville, AL, USA}}
\newcommand{\uofa}{\affiliation{Lunar and Planetary Laboratory, University of Arizona, Tucson, AZ 85721, USA}}

\graphicspath{{./}{figures/}}

\received{}
\revised{}
\accepted{2019/11/20}

\submitjournal{ApJS - Parker Solar Probe Special Issue}

\shorttitle{The Solar Probe Cup}
\shortauthors{Case et al.}
\begin{document}
\title{The Solar Probe Cup on Parker Solar Probe}

\correspondingauthor{Anthony W. Case}
\email{tonycase@cfa.harvard.edu}

\author[0000-0002-3520-4041]{A. W. Case}\sao
\author[0000-0002-7077-930X]{Justin C. Kasper}\umich\sao
\author[0000-0002-7728-0085]{Michael L. Stevens}\sao
\author[0000-0001-6095-2490]{Kelly E. Korreck}\sao
\author[0000-0002-5699-090X]{Kristoff Paulson}\sao
\author{Peter Daigneau}\sao
\author{Dave Caldwell}\sao
\author[0000-0002-9821-9662]{Mark Freeman}\sao
\author[0000-0002-7628-6071]{Thayne Henry}\csdl
\author{Brianna Klingensmith}\csdl
\author{Miles Robinson}\ucb
\author{Peter Berg}\ucb
\author{Chris Tiu}\gsfc
\author[0000-0003-1940-0605]{K. H. Wright, Jr.}\usra
\author{David Curtis}\ucb
\author[0000-0001-5051-1536]{\added{Michael Ludlam}}\ucb
\author[0000-0001-5030-6030]{Davin Larson}\ucb
\author[0000-0002-7287-5098]{Phyllis Whittlesey}\ucb
\author[0000-0002-0396-0547]{Roberto Livi}\ucb
\author[0000-0001-6038-1923]{Kristopher G. Klein}\uofa
\author[0000-0002-7365-0472]{Mihailo M. Martinovi\'c}\uofa

\begin{abstract}
The Solar Probe Cup (SPC) is a Faraday Cup instrument onboard NASA's Parker Solar Probe (PSP) spacecraft designed to make rapid measurements of thermal coronal and solar wind plasma.    The spacecraft is in a heliocentric orbit that takes it closer to the Sun than any previous spacecraft, allowing measurements to be made where the coronal and solar wind plasma is being heated and accelerated.  The SPC instrument was designed to be pointed directly at the Sun at all times, allowing the solar wind (which is flowing primarily radially away from the Sun) to be measured throughout the orbit.  The instrument is capable of measuring solar wind ions with an energy/charge between 100 V and 6000 V (protons with speeds from $139-1072~km~s^{-1})$.  It also measures electrons with an energy between 100 V and 1500 V.  SPC has been designed to have a wide dynamic range that is capable of measuring protons and alpha particles at the closest perihelion (9.86 solar radii from the center of the Sun) and out to 0.25 AU.  Initial observations from the first orbit of PSP indicate that the instrument is functioning well.

\end{abstract}

%%%%%%%%%%%%%%%%%%%%%%%%%%%%%%%%%%%%%%%%%%%%%%%%%%%%%%%
\section{Introduction} \label{sec:intro}
\subsection{Spacecraft and Suite}
Parker Solar Probe (PSP) is a robotic NASA mission that consists of a single three-axis-stabilized (primarily Sun-pointed) spacecraft in a heliocentric orbit with an aphelion beyond the orbit of Venus and a perihelion that gradually decreases from 35 solar radii (R$_S$) to 9.86 R$_S$ over the course of 24 orbits and seven years through the use of seven Venus gravity assists.  The primary objectives of the mission are described in detail in \citet{fox_2016} and are briefly summarized here: (1) Determine the structure and dynamics of the magnetic fields at the sources of the fast and slow solar wind.  (2) Trace the flow of energy that heats the solar corona and accelerates the solar wind.  (3) Explore mechanisms that accelerate and transport energetic particles.

To fully address all of these science objectives, it is necessary to measure the thermal plasma present in the solar corona and solar wind.  The ``Solar Wind Electrons, Alphas, and Protons'' (SWEAP) investigation is a suite of particle-sensing instruments that are a part of the Parker Solar Probe (PSP) payload.  The overarching objectives and measurement concepts of the suite are described in detail by \citet{kasper_2016}.  The suite consists of 3 ``Solar Probe ANalyzer'' (SPAN) instruments and the Solar Probe Cup (SPC).  The SPAN instruments are electrostatic-analyzers (ESAs), which are capable of providing the full three-dimensional velocity distribution function (VDF) through the use of curved plates for discrimination based on the incoming particle's energy/charge, multiple anodes to measure the flux from different azimuth angles, and electrostatic deflectors to scan through elevation angles \citep{whittlesey_2019, livi_2019}.  Additionally, the SPAN-Ai instrument contains a time-of-flight (ToF) section that allows identification of a particle's mass.  The SPAN-Ai instrument is blocked from the Sun by the spacecraft's heat shield, but it is anticipated that near the Sun, especially later in the mission when the orbital motion of the spacecraft is large, solar wind ions will flow at a large angle from radial and may be seen by SPAN-Ai.  The rest of the time the core of solar wind ions cannot be observed by an instrument behind the heat shield of the spacecraft.  SPC is primarily designed to fill this observational gap by pointing at the Sun all the time.  SPC is a Faraday-cup instrument, the in-depth description of which will follow in the remainder of this article.  The four instruments within the suite are briefly summarized in Table \ref{tab:sweap_instruments}.

\startlongtable
\begin{deluxetable*}{c c c c c}
\tablecaption{The instruments within the SWEAP suite \label{tab:sweap_instruments}}
\tablehead{
\colhead{Name} & \colhead{Type} & \colhead{Particle Measured} & \colhead{Measurement Type} & \colhead{Look Direction}
}
\startdata
SPAN-Ai & Electrostatic Analyzer + ToF	& Ions		                & 3D VDF + mass      & Ram \\
SPAN-Ae & Electrostatic Analyzer 	& Electrons		            & 3D VDF       & Ram \\
SPAN-Be & Electrostatic Analyzer 	& Electrons		            & 3D VDF     & Anti-Ram \\
SPC & Faraday Cup                   & Ions and Electrons		& 1D VDF + energy-dependent flow angles   & Nadir
\enddata
\end{deluxetable*}

\subsection{The Solar Probe Cup}
The Solar Probe Cup (SPC) is a Faraday cup instrument that is designed to measure the ions and electrons that make up the solar wind and coronal plasma.  Previous generations of these types of instruments have flown successfully on numerous missions including \edit1{the Voyager 1 and 2 plasma science experiment (PLS)} \citep{bridge_1977}, Wind \citep{ogilvie_1995}, Spektr-R \citep{zastenker_2013}, and the Deep-Space Climate Observatory (DSCOVR). Previous iterations of the SPC design concept were reported in \citet{case_2013} (before the instrument's preliminary design review) and \citet{kasper_2016} (before the instrument's critical design review), but the literature has thus far not captured the final design, which continued to evolve up to and beyond the critical design review.  The rest of this article provides information about the instrument design and operation with the objective of being a useful introduction and guide to the instrument for prospective users of Solar Probe Cup data.

Since the solar wind flows primarily radially away from the Sun, it is desirable to have an instrument pointed directly at the Sun for those times when the spacecraft's azimuthal velocity is small relative to the solar wind speed.  The desire to be pointed directly at the solar wind results in the instrument also being exposed directly to the solar photon flux, which can be as much as 475 times higher than the flux seen at 1 AU.  Thus, while SPC's measurement method is extremely similar to previous instruments such as the Sun-pointing Voyager PLS and DSCOVR Faraday Cup, its material construction is vastly different.

The instrument consists of a set of parallel, planar metal grids that produce the necessary electrostatic fields to modulate the flow of particles based on their energy/charge.  The grids and housing are constructed out of refractory metals and alloys (e.g., tungsten, molybdenum, niobium) and sapphire insulators to resist the high temperatures encountered near the Sun. The electric field produced by a grid with oscillating high-voltage potential sorts particles based on their energy/charge.  Metal plates collect the charge from the incident charged particles and that current is sent to an electronics board that amplifies and digitizes the oscillating portion of the signal.  A ``synchronous detection'' is then performed to detect the amplitude of the signal that occurs at the same frequency as the time-varying potential on the high-voltage grid.  \edit1{In contrast to previous experiments, the synchronous detection on SPC is performed in digital electronics, rather than through the use of an analog demodulation technique.}

An FPGA (Field-Programmable Gate Array) commands the instrument to scan through a series of different high-voltage modulator voltage waveforms.  The DC voltage of the waveform determines the center of the voltage window, and the AC portion determines the \deleted{voltage (and thus }energy/charge\deleted{)} resolution.  The primary data product produced by the instrument is a measurement of the current in a set of energy/charge windows that make up a spectrum.  These current spectra can then be transformed into a one-dimensional velocity distribution function.  Further data processing is able to produce fluid parameters such as velocity, density, and temperature by taking moments of, or fitting curves to, the velocity distribution function.

The rest of the paper is organized as follows: Section \ref{sec:instrument_description} describes the mechanical design of the instrument and Section \ref{sec:electronics} describes the measurement electronics.  Section \ref{sec:instrument_operation} discusses how the instrument is operated throughout a typical Parker Solar Probe orbit.  Section \ref{sec:data_description} describes the data products that will be available to the public.  Section \ref{sec:observed_performance} provides some example measurements made during the commissioning period and first orbit of the spacecraft to demonstrate the on-orbit performance of the instrument.

%%%%%%%%%%%%%%%%%%%%%%%%%%%%%%%%%%%%%%%%%%%%%%%%%%%%%%%
\section{Instrument Description} \label{sec:instrument_description}
The operating principle of the Solar Probe Cup is extremely similar to that of previous Faraday cups like Wind/SWE \citep{ogilvie_1995} and the Deep-space Climate Observatory (DSCOVR).  In particular, the DSCOVR Faraday cup, as with SPC, is Sun-pointed with a segmented collector plate to allow for the determination of the flow-direction of the incoming particles.  The primary changes to SPC from those previous instruments are its size (much smaller due to the high particle fluxes closer to the Sun) and the materials from which it is made (to withstand the high temperatures in the near-Sun environment.)

%%%%%%%%%%%%%%%%%%%%%%%%%
\subsection{Instrument Mechanical Design}
\begin{figure*}[ht!]
\plotone{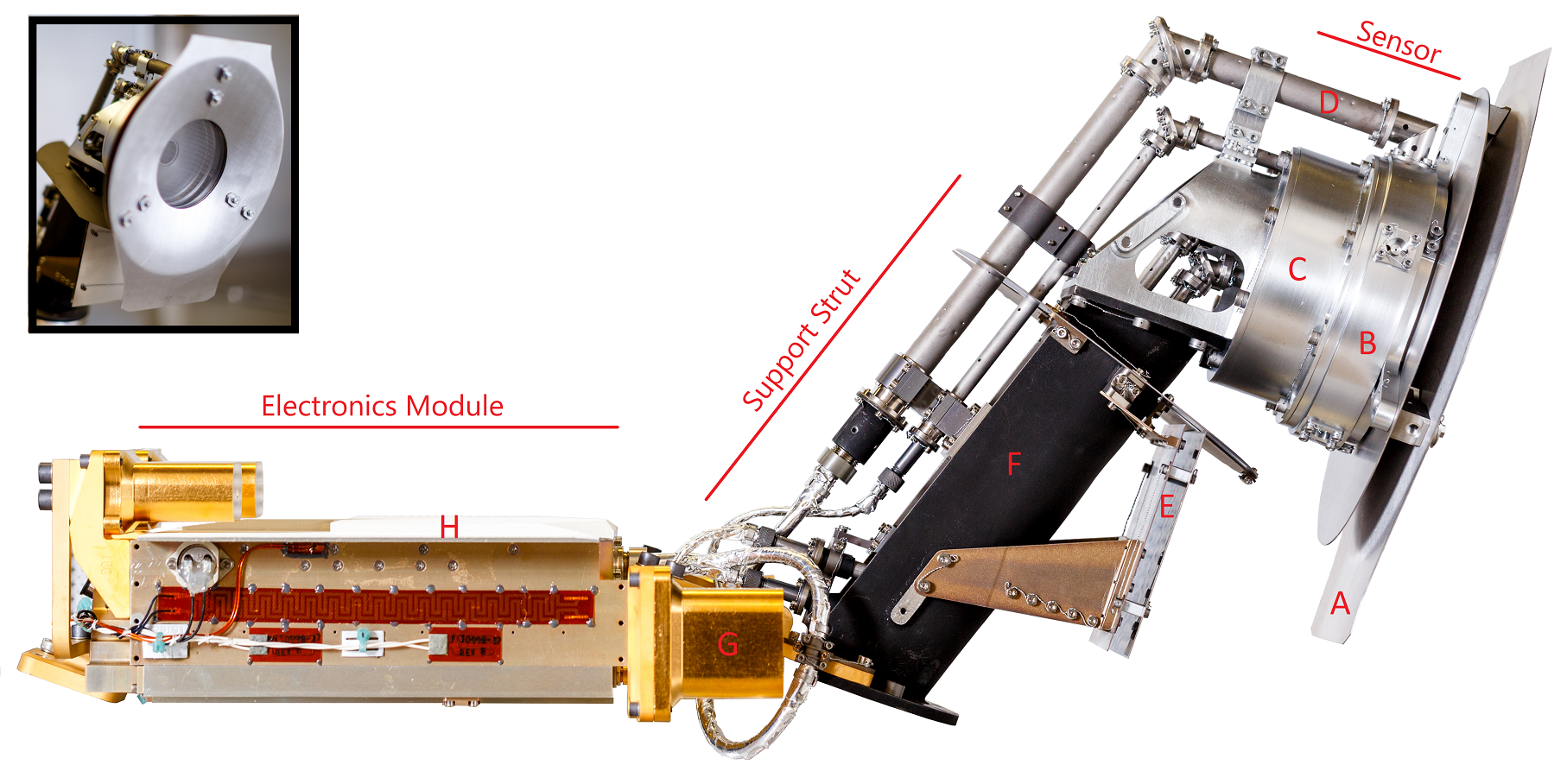}
\caption{The Solar Probe Cup instrument.  The right-hand-side faces toward the Sun. Notable features are labeled: \textbf{A:} thermal shield, \textbf{B:} high-voltage modulator subassembly, \textbf{C:} collector plate subassembly, \textbf{D:} custom high-voltage coaxial cable, \textbf{E:} secondary thermal shield, \textbf{F:} support strut, \textbf{G:} vibration dampening mechanism, \textbf{H:} electronics box radiator.  \textbf{Inset:} Front-view of the SPC sensor. \label{fig:spc_profile}}
\end{figure*}

Figure \ref{fig:spc_profile} shows a side-view of the flight version of the instrument.  The right-hand-side (as shown in Figure \ref{fig:spc_profile}) faces the Sun during nominal science operations.  The instrument is made up of two major subassemblies: the sensor and the electronics module.  The sensor is held in place by the ``support strut'', which positions the instrument outside of the shadow cast by the spacecraft thermal shield.  The front of the sensor is directly impacted by light and particles emanating from the direction of the Sun.  The front of the electronics module sits about 26 centimeters away from the back of the sensor so that all of the electronics are within the shadow of the spacecraft thermal shield and operate within more typical temperatures required of electronic components.

Important aspects of the instrument are labeled in Figure \ref{fig:spc_profile}.  The thermal shield (``A'') reflects, absorbs, and re-radiates any solar photon flux that would have otherwise impacted the instrument.  The thermal shield has an open aperture in its center that allows particles to pass into the sensitive portion of the instrument (see inset front-view in Figure \ref{fig:spc_profile}).  Particles making it through that aperture enter the high-voltage (HV) modulator subassembly (``B''), where the E/q selection takes place.

\begin{figure*}[ht!]
\plotone{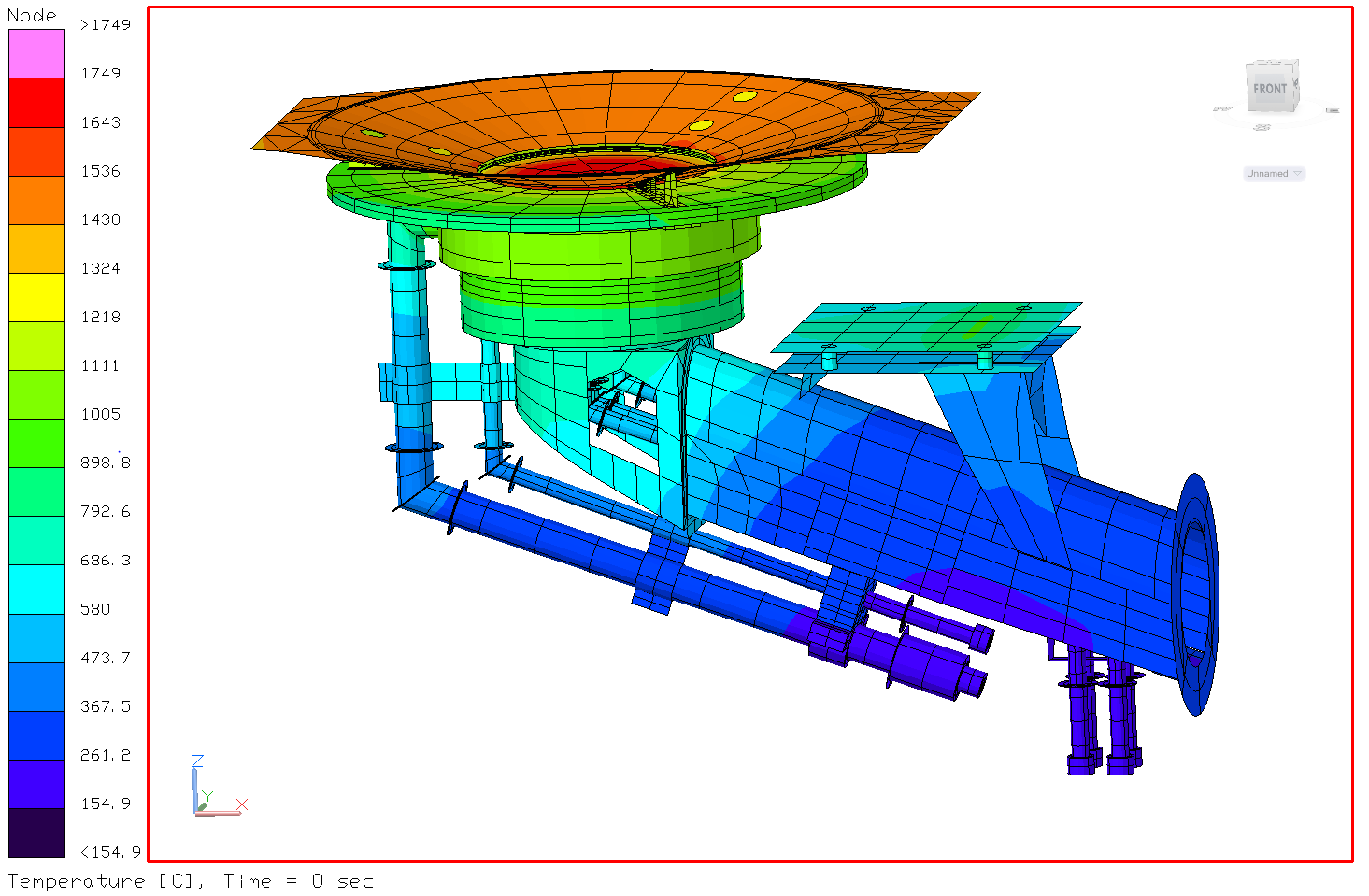}
\caption{Expected temperatures of the various SPC components at closest approach to the Sun. \label{fig:spc_thermal}}
\end{figure*}

The expected temperature distribution for the final SPC sensor design at closest approach (9.86 Rs) is shown in Figure \ref{fig:spc_thermal}.  The thermal modeling was performed in Thermal Desktop.  Temperature-variant properties, both thermo-physical and thermo-optical, were employed for the major materials in the sensor design.  High-temperature thermo-optical properties (absorptance and emittance) were derived from sample testing at the PROMES solar furnace in France \citep{brodu_2014, brodu_2015}, as well as through model correlations performed with a qualification model SPC sensor in the Solar Environment Simulator (SES) \citep{cheimets_2013}, a test chamber developed specifically for the testing of this instrument.

%%%%%%%%%%%%%%%%%%%%%%%%%
\subsection{Electrostatic Optics Design}

The instrument uses a retarding electric field to either accept or reject incoming particles based on their energy per charge.  A series of planar and parallel metal grids in the modulator and collector subassemblies are used to control the potential in the sensor.  The metal grids have a transparency of $\sim$0.9, allowing the majority of particles to flow through them unimpeded.  The electric field is created by controlling the voltage on a high-voltage (HV) grid that is surrounded by a grid on either side that is at ground potential.

When a positive potential is placed onto the HV grid, the electrical field repels positively-charged particles.  If a particle has sufficient energy/charge, it is slowed, but not repelled by the electric field and eventually makes its way through the HV grid.  On the other side of the HV grid, the electric field points in the opposite direction and re-accelerates the particle to its original speed.

The collector subassembly (``C'') is where the charge from the incoming charged particles is deposited.  The charge collection occurs in the ``collector plates'', which are metal plates (made of niobium) in the shape of the 4 quadrants of a circle, with each plate isolated from the others and from the rest of the instrument housing using a sapphire substrate.  The signal from each of these collector plates is transmitted down its own coaxial cable to the electronics module.

A ``suppressor grid'' is placed directly above the collector plates to aid in the collecting of charge from the incoming particles.  A negative voltage (-55 V) is placed onto the suppressor grid, so that any secondary electrons, ejected from the collector plate due to the impact of the incoming primary particle, are repelled back toward the collector plate.  \edit1{Less than 1\%  of those secondary electrons have energies greater than 55 V, and any additional signal from escaping secondaries is corrected for with the in-flight calibration.  Additionally, the suppressor grid repels secondaries that are ejected from the collector plates due to the large number of photons that impinge on the collector plates. Without that feature the measurement circuitry would be required to source more current to keep the collector plates near ground potential.}

\begin{figure*}[ht!]
\plotone{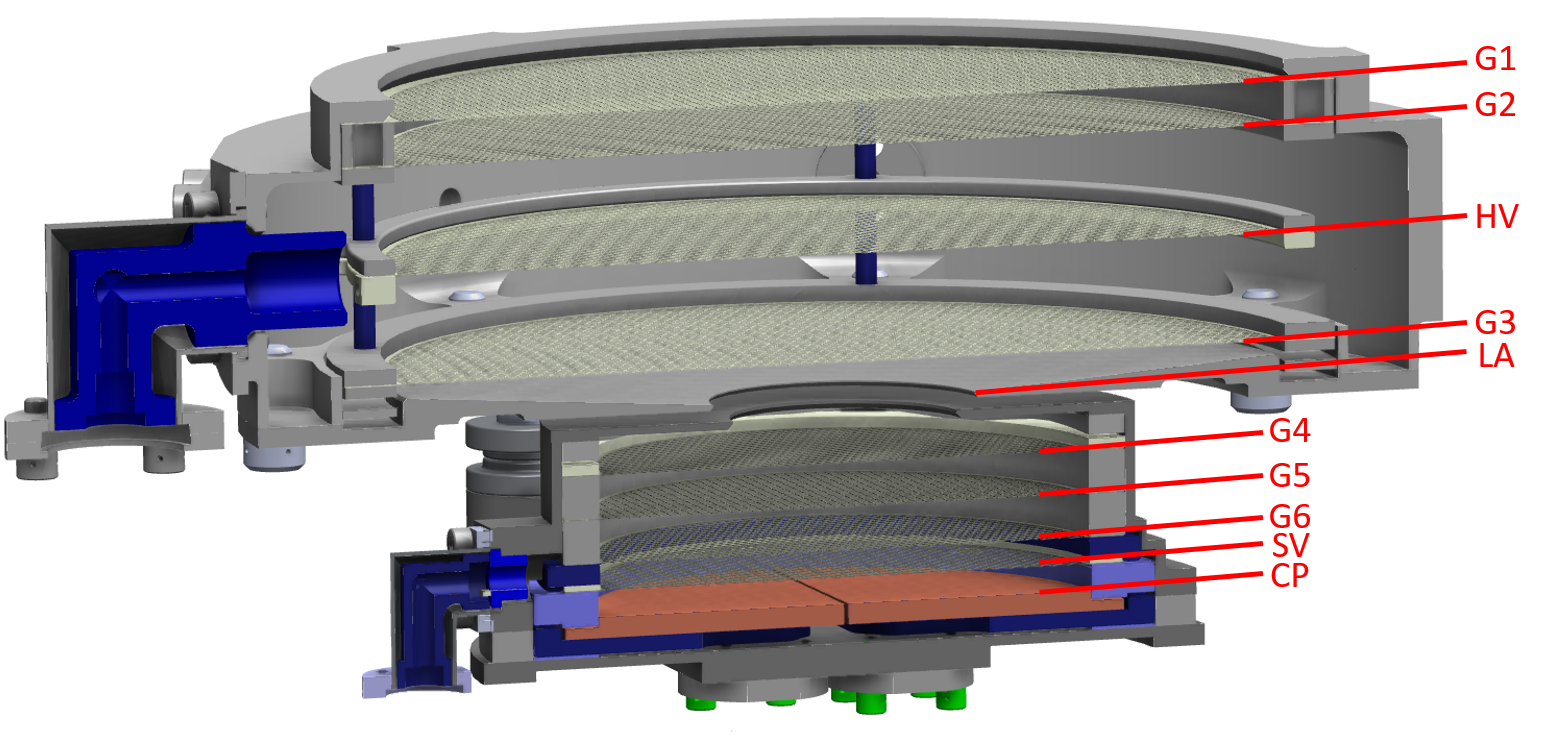}
\caption{Cross-section of a computer drawing of the Solar Probe Cup instrument.  The top of the image is the Sun-facing side.  The thermal shield\edit1{, which defines the entrance aperture (EA), }is not shown \edit1{in this view}.  Blue components are made of sapphire (an electrical insulator).  The other components are made of various metals (tungsten, niobium, molybdenum, and alloys of those metals including titanium, hafnium, zirconium, and other trace elements). \edit1{The labels refer to each component's acronym in the first column of Table \ref{tab:spc_sizes_locs}} \label{fig:spc_section_oblique}}
\end{figure*}

Figure \ref{fig:spc_section_oblique} shows a cross-section of the modulator and collector subassemblies.  Each planar feature (aperture, grid, or collector plate) has been labeled with a two-letter identifier that corresponds to the first column in Table \ref{tab:spc_sizes_locs}, where the precise locations and sizes of each component are listed.  \edit1{The entrance aperture (EA) is defined by the inner edge of the SPC thermal shield, which is not shown in Figure \ref{fig:spc_section_oblique}}.  The ground grid ``G1'' is placed at the top of the instrument so as to shield the oscillating electric field from reaching any area outside of the instrument.  The ground grids ``G4'', ``G5'', and ``G6'' are placed between the modulator and the collector plates to reduce the capacitive coupling between the modulating voltage on the ``HV'' grid and the collector plates.

\startlongtable
\begin{deluxetable*}{c c|c c c}
\tablecaption{Locations and sizes of electrostatic optics elements within SPC \label{tab:spc_sizes_locs}}
\tablehead{
\colhead{Identifier} & \colhead{Description} & \colhead{Radius\tablenotemark{a}} & \colhead{Axial\tablenotemark{b}} & \colhead{Thickness} \\
\colhead{} & \colhead{} & \colhead{(mm)} & \colhead{(mm)} & \colhead{(mm)}
}
\startdata
EA & Entrance Aperture			& 39.9		& 50.3 & n/a \\
G1 & Modulator Ground Grid		& 42.46	    & 45.2 & 0.1 \\
G2 & Modulator Ground Grid     	& 42.46 	& 41.1 & 0.1 \\
HV & High-voltage Grid			& 42.47		& 31.0 & 0.1 \\
G3 & Modulator Ground Grid     	& 42.46 	& 20.9 & 0.1 \\
LA & Limiting Aperture			& 10.86  	& 16.9 & n/a \\
G4 & Collector Ground Grid    	& 23.79		& 12.8 & 0.1 \\
G5 & Collector Ground Grid      & 23.79		& 9.45 & 0.1 \\
G6 & Collector Ground Grid      & 23.79		& 6.1  & 0.1 \\
SV & -55 VDC Suppressor Grid	& 23.79		& 3.0  & 0.1 \\
CP & Collector Plates			& 23.94  	& 0.0  & n/a
\enddata
\tablenotetext{a}{For grids, this is the radius of the transparent portion of the grid}
\tablenotetext{b}{Axial distance from the top of the collector plates to the bottom of the component}
\end{deluxetable*}

%%%%%%%%%%%%%%%%%%%%%%%%%%%%%%%%%%%%%%%%%%%%%%%%%%%%%%%%%%%%%%%%%%%%%%%%%%%
\section{Electronics}
\label{sec:electronics}

\begin{figure}[ht!]
\plotone{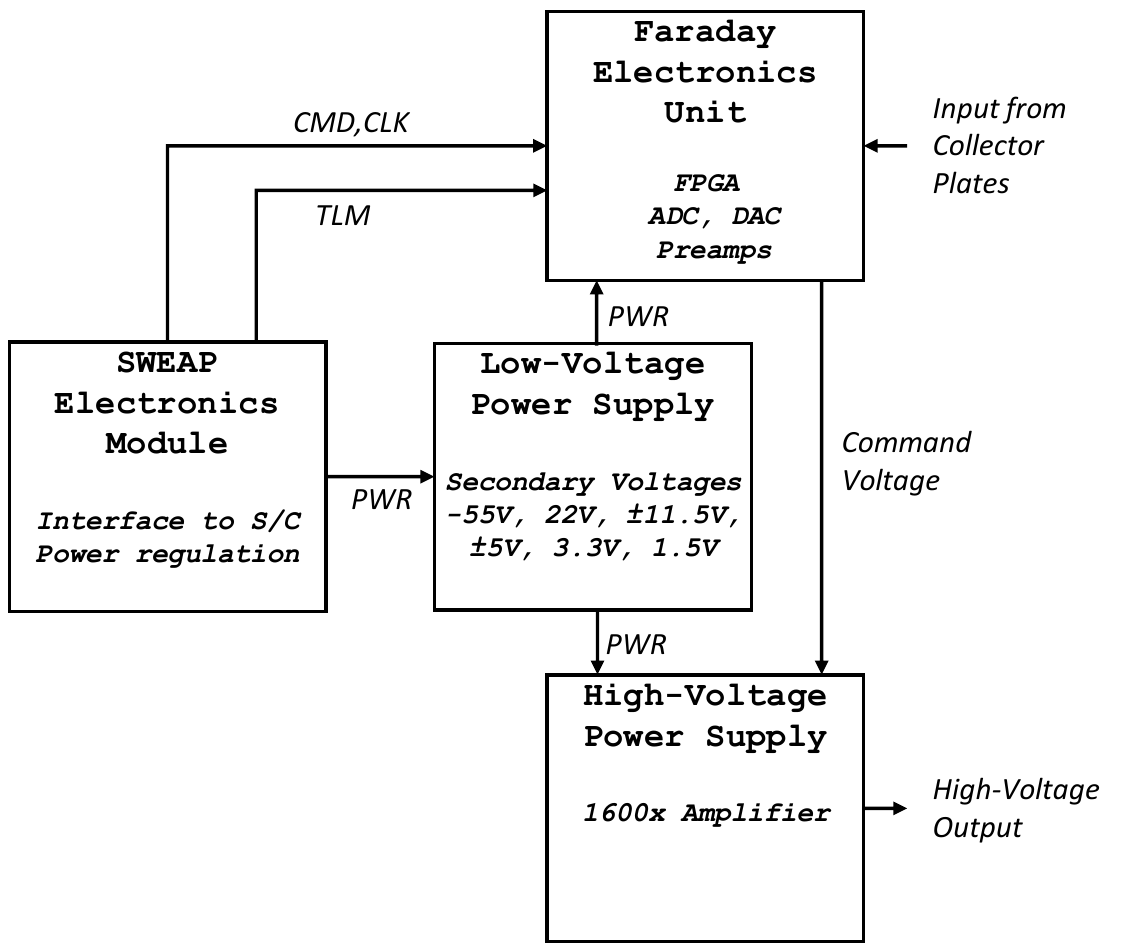}
\caption{Electronics Module Block Diagram \label{fig:instrument_block}.  The three SPC electronics boards are shown along with the signals that flow between them and the suite processing unit.}
\end{figure}

The ``SWEAP Electronics Module'' (SWEM) is the suite processing unit that stores SPC`s command sequences, operations tables, science configuration parameters, and recorded data.  The SWEM is the sole SWEAP interface to the spacecraft.  The block diagram in Figure \ref{fig:instrument_block} shows the interconnection of the SWEM and the three electronics boards within SPC.  The function of those boards is as follows: 1) The Faraday Electronics Unit (FEU) board has measurement electronics which amplify and digitize the incoming signals from the sensor and an FPGA for signal processing and communication with the SWEM, and is described in Section \ref{sec:measurement_electronics}.  2) The High-Voltage Power Supply (HVPS) is a 1600:1 amplifier that receives a control voltage from the FEU board, amplifies that signal, and places the voltage onto the high-voltage modulator grid inside the sensor, and is discussed in Section \ref{sec:hvps}.  3) The Low-Voltage Power Supply (LVPS) receives regulated input power and provides secondary voltages to the FEU and HVPS as well as providing -55 V to the suppressor grid in the collector subassembly.

\subsection{Measurement Electronics}\label{sec:measurement_electronics}
The primary objective of the measurement electronics is to amplify the signal coming from each collector plate and measure the amplitude of the modulated waveform at the synchronous detection frequency.  As opposed to previous Faraday cups designed for solar wind measurements, much of the data processing is performed in digital electronics instead of analog electronics (as was done on DSCOVR, Wind, and Voyager \citep{bridge_1977, ogilvie_1995}.

\begin{figure*}[ht!]
\plotone{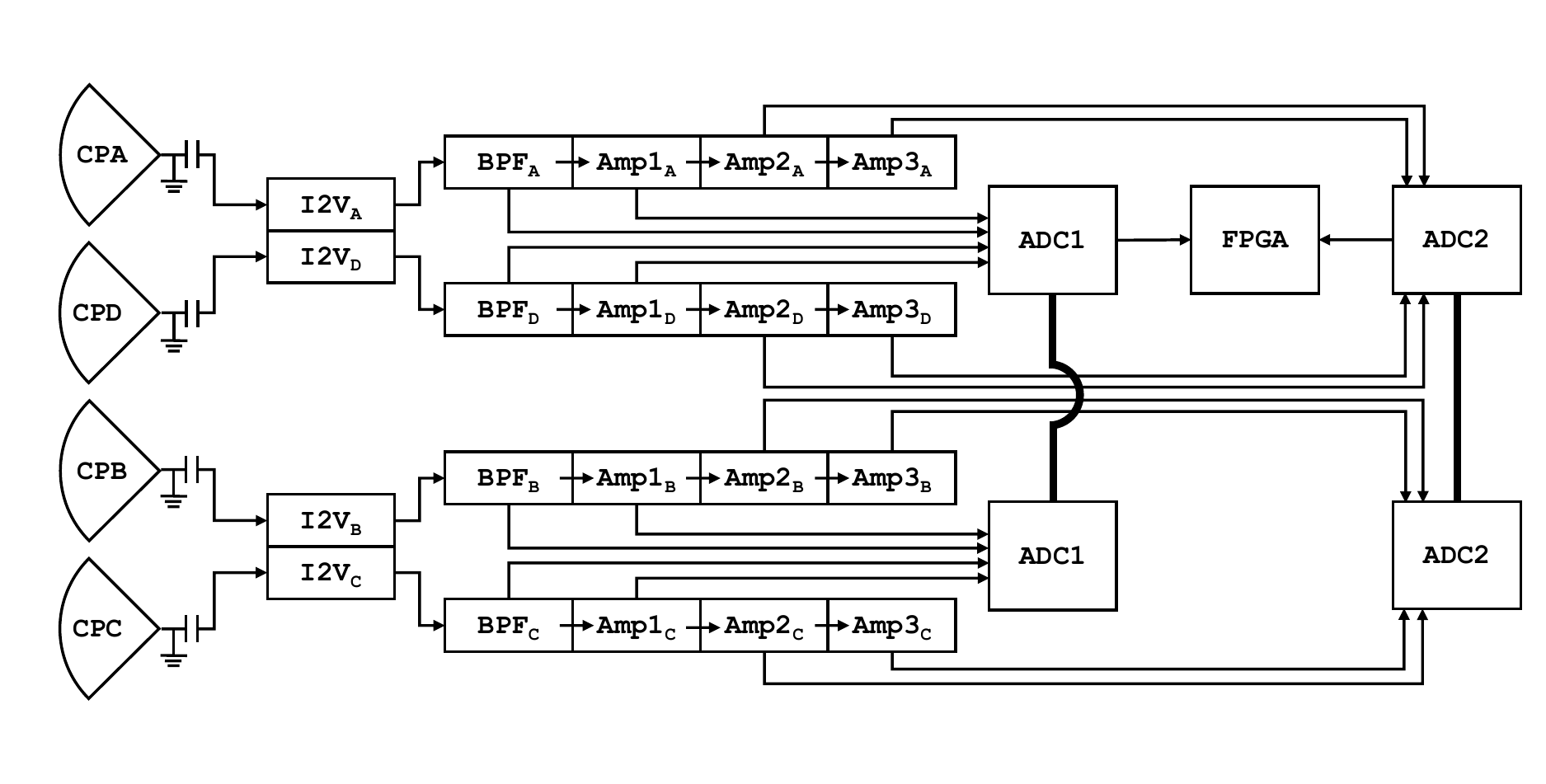}
\caption{SPC Electronics Block Diagram.  Each outlined shape represents an electronic circuit; shapes that are touching one another indicate circuits whose major components are all part of a multiple circuit IC package. ADC1 and ADC2 \added{are} each a single chip, but separated in the diagram for ease of display. CP=Collector Plate. I2V=Current-to-Voltage Amplifier. BPF=Bandpass Filter. Amp=Amplifier. ADC=Analog-to-Digital Converter. FPGA=Field Programmable Gate Array. \label{fig:spc_elec_block}}
\end{figure*}

Figure \ref{fig:spc_elec_block} shows a block diagram of the electronics circuit used to accomplish this task.  The various stages of the measurement circuit are described below from left to right as shown in Figure \ref{fig:spc_elec_block}.  The current waveform that is present on the signal lines from each collector plate is AC-coupled so that only the oscillating portion of the current is passed into the input amplifier.  The AC current is converted to a voltage waveform through an ADA4610-2S op-amp configured as a trans-impedance amplifier, which is then amplified and filtered through a bandpass filter centered on the modulation frequency (using an OP484S series op-amp).  The filter has a bandwidth of about $\pm 215$ Hz.  Three more amplification stages further amplify the signal: the first amplification stage uses the output of the bandpass filter as its input, and each further amplification stage uses the output of the previous stage as its input with further AC coupling to remove offsets. 

The output from the bandpass filter and each of the 3 amplification stages are fed into a multiplexed ADC (ADC128S102) where they are digitized at 37.5 kHz.  The bandpass filter and first amplification stages from all collector plates are fed into a single ADC chip with 8 inputs.  The outputs from the second and third amplifications stages are fed into a separate ADC.  \edit1{Each gain stage amplifies the signal by approximately a factor of 16.  Pre-flight simulations suggest that only the three stages with the highest gains will be used in the expected solar wind conditions and operating configurations.}

Thus, there are digitized waveforms from 4 different gains from each of the 4 collector plates (16 signals total) that enter the FPGA.  Each signal consists of an oscillating waveform that is highly filtered so that it contains primarily the modulation frequency.  The FPGA receives the values from each of the ADCs and performs digital processing to calculate the amplitude of the waveform for components both in-phase and out-of-phase with the high-voltage modulation waveform.

The FPGA and its associated firmware (which will be referred to as simply the FPGA) has three essential tasks.  1) To communicate with the suite processor unit (SWEM) and its associated flight software for purposes of receiving commands and sending telemetry.  2) To analyze the incoming waveforms from the measurement electronics.  3) To decide which voltage windows to measure and to command the high-voltage power supply to produce the necessary high-voltage waveforms to send to the high-voltage modulator grid.  Task number 1 will be more thoroughly discussed in Section \ref{sec:instrument_operation}, while Section \ref{sec:single_measurement} will continue with a description of the analysis of the signal waveforms in the FPGA and high-voltage power supply commanding.

%%%%%%%%%%%%%%%%%%%%%%%%%
\subsection{Measurement within a Single Voltage Window} \label{sec:single_measurement}
All measurements in SPC occur in a ``measurement window'' during which a sinusoidal high-voltage waveform is commanded, and the currents that fall onto each of the four collector plates are measured.  The length of this window (the ``measurement time'', $MT$) can be commanded to any time $2^n/1171.875\: Hz$, where n is an integer and $2\leq n\leq 8$.  

\begin{figure*}[ht!]
\plotone{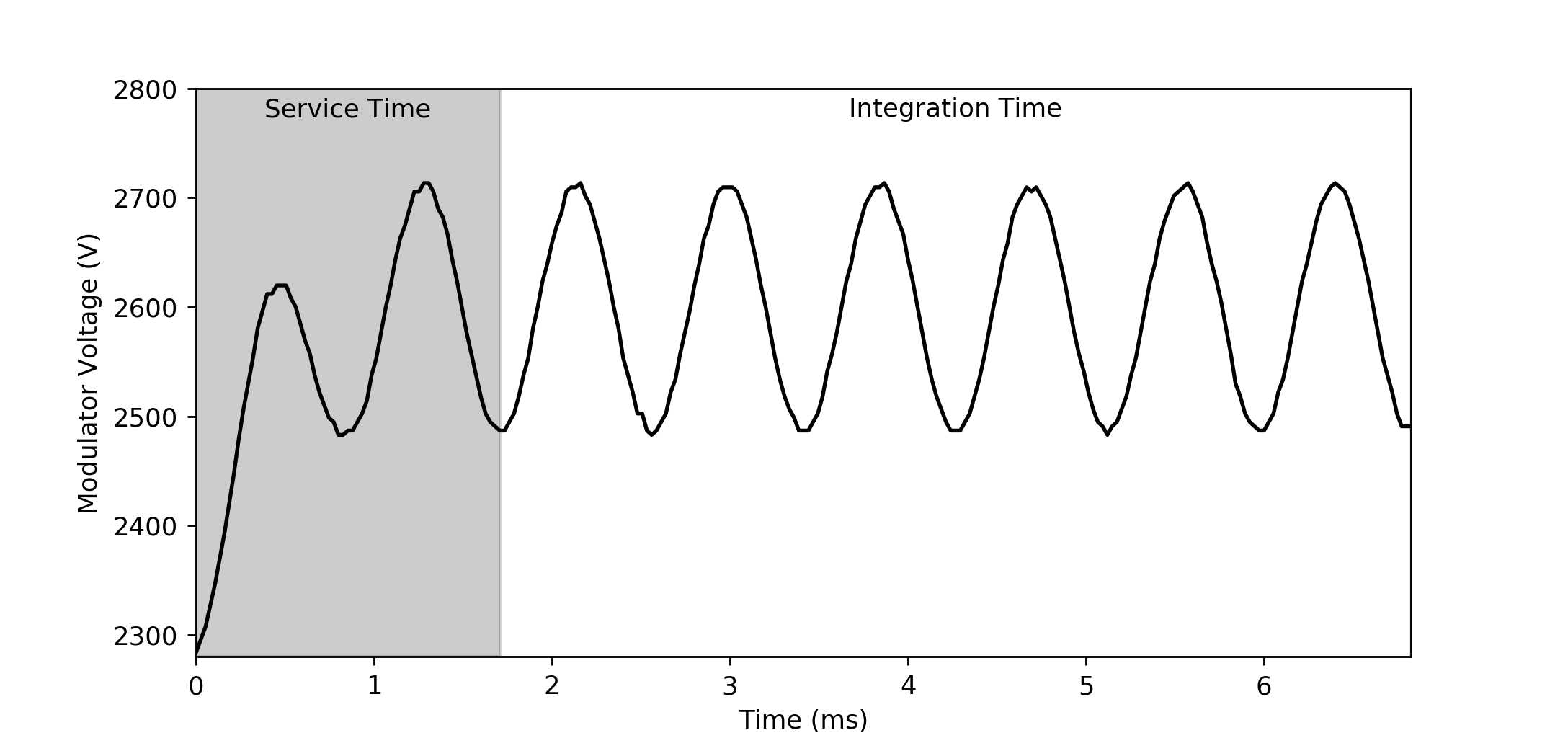}
\caption{Commanded high-voltage over an SPC measurement period for operation parameters typically used during a solar encounter period.  During the Service Time, the high-voltage waveform is changing to the newly-commanded voltage and stabilizing.  During the Integration Time, data are accumulated in the measurement circuitry. \label{fig:measurement_time}}
\end{figure*}

This measurement time is broken into two distinct periods.  1) The ``service time'', $ST$, during which the modulation voltage waveform is being changed to the newly requested waveform, and 2) the ``integration time'', $IT$, during which measurements from each collector plate are accumulated.  The firmware implementation requires $ST=m/1171.875\: Hz$ with $1\leq m\leq 7$ and $IT=MT-ST$.  As an example, typical SPC operation during an encounter uses $n=3, m=2$, so that $MT=6.82\overline{6}\: ms$, $ST=1.70\overline{6}\: ms$, and $IT=5.12\: ms$.  Figure \ref{fig:measurement_time} shows the voltage being supplied by the high-voltage power supply for a relatively high voltage window (so that the housekeeping circuit noise is minimized) over the course of a single measurement time with the aforementioned ST and IT.

During the integration time the amplitude of the waveform must be determined for each of the 16 signals coming from the measurement electronics.  This is accomplished through what is essentially a Fourier transform performed at a single frequency (the modulation frequency of 1171.875 Hz). The FPGA takes each sample from the ADC and multiplies it by both a sine and a cosine reference waveform.  It then keeps a running sum of all sine and all cosine values throughout the integration time.  At the end of the integration time the sums are divided by the number of samples measured and then scaled to fit within a 12-bit value.

That analysis process results in 32 12-bit values (a sine and a cosine amplitude for each of the four amplification stages from each of the four collector plates), but only 8 values are actually put into the telemetry packet (a sine and cosine value for each collector plate).  For each collector plate, the FPGA selects the highest gain stage that is not saturated and puts those values into the telemetry packet that gets sent to the SWEM for further processing.

%%%%%%%%%%%%%%%%%%%%%%%%%
\subsection{High-Voltage Power Supply}\label{sec:hvps}
The High-Voltage Power Supply (HVPS) receives a control signal from the FEU that varies between -0.93 and 3.75 volts, resulting in outputs between about -1500 and +6000 volts (the negative voltages are used only rarely, for electron measurements.)  The HVPS functions as a voltage amplifier with a gain of 1600 to convert the control signal into a high voltage output driving the high-voltage modulator grid. The supply is designed to respond to control signals that vary at 1.2kHz or slower and is capable of sourcing up to 100 microamps of current. It achieves this by generating two high voltage DC ``rails'', one at +6 kV and another at -1.5 kV.  A feedback network compares the output to the control signal, and differences cause either the ``pull-up'' or ``pull-down'' high-voltage chains to allow current to flow between one of the rails and the output. This is achieved by causing current to flow through one of two banks of transformers. One bank allows current to pass through a set of nine transistors connecting the positive rail to the output, while the other drives a set of transistors connecting the negative rail to the output. The transformers act as a barrier between the low-voltage control circuitry and the high voltage output.

Though the HVPS is intended to simply amplify every input voltage by a factor of 1600, in reality it has some features that affect the shape and amplitude of the output waveform produced.  The control waveform provided to the HVPS is always a nearly perfect sinusoid whose amplitude is determined based on the user's desired energy resolution.  When the HVPS board is at high temperatures, an amplified version of the control waveform is accurately reproduced.  However, at lower temperatures the HVPS has trouble reproducing the full amplitude of the requested waveform.  When the requested peak-to-peak amplitude of the output is below $\sim$1000 volts, the output amplitude can be as small as 30\% of the requested amplitude.  \edit1{The amplitude of the HV waveform is not measured during operation, so corrections must use pre-flight test data and in-flight comparisons with FIELDS electron density measurements to compensate.}  This undershoot is accounted for in two ways: 1) artificially large amplitude waveforms are requested so that the output amplitude is closer to our actual desired amplitude.  2) The data processing algorithms account for the fact that a narrower region of energy/charge space is being measured; all distributed data have had this correction applied.

%%%%%%%%%%%%%%%%%%%%%%%%%%%%%%%%%%%%%%%%%%%%%%%%%%%%%%%
\section{Instrument Operation} \label{sec:instrument_operation}
The primary interface to the SPC instrument is through the SWEM.  The SWEM is able to store command sequences that initialize SPC into an operational configuration.  First, the SWEM transfers to SPC the high-voltage tables that define the voltage range and resolution that should be used in each measurement scan.  Then the rest of the configuration parameters (e.g., integration time, and service time) are uploaded.  Included in those parameters are commands that tell SPC whether to turn on its internal calibration source, or to go into electron or flux-angle mode.   After all parameters are uploaded, the SWEM can command SPC to start making measurements.  After that start command, all facets of operation are performed within the SPC FPGA until a stop command is received from the SWEM.

%%%%%%%%%%%%%%%%%%%%%%%%%%%
\subsection{Scanning through a Range of Voltage Windows}
SPC operates by starting at a low voltage, making a measurement of a single voltage window, then stepping up to a slightly higher voltage and making another measurement.  It continues this process until it reaches the maximum voltage that has been commanded, at which point it ramps the high-voltage power supply down to the voltage of the beginning of the next scan.  The FPGA prefers to use exactly one measurement time to ramp the high-voltage supply down to the voltage for the first window in the next scan, but it never uses less than $6.28\overline{6}\: ms$.

The instrument can be programmed to autonomously switch between two different voltage window scanning modes.  The first, called ``full-scans'', consists of a scan through the entire range of voltages measured by the instrument.  Typically, this consists of very wide voltage windows, so that the entire voltage range can be covered in a relatively short amount of time.  The purpose of the full-scans is to quickly identify the voltage at which the peak current is measured.  With that knowledge future scans can focus in on a narrower voltage range.  

``Peak-tracking'' scans allow the instrument to improve its voltage resolution and/or its measurement cadence by only measuring a range of voltages that are nearby the peak of the velocity distribution function.  When commanded to start taking measurements, SPC always begins with a full-scan.  During the course of a full-scan, the SPC FPGA keeps track of the location of the highest current that was measured.  A number of configurable parameters determine the next action taken by the FPGA.  (1) Peak Repeat Count: The instrument only performs a set number of peak-tracking scans before going back and doing another full-scan.  This is to ensure that a ``false peak'' (perhaps caused by a single noisy measurement) has not inadvertently caused the instrument to be scanning the wrong voltage range.  (2) Magnitude of the Current Measured: If a peak signal is not strong enough, then the peak-tracking algorithm will perform another full-scan rather than performing a peak-tracking scan around a weak signal.  (3) Proximity to Voltage Extrema: If the peak signal is too near the lower or upper boundary of the voltage range being measured, the peak-tracking algorithm will perform a full-scan rather than a peak-tracking scan in case the actual peak of the VDF is somewhere outside of the measurable voltage range.

Because of the wider voltage windows typically used during full-scans, it is common that the VDF measured during those scans will differ from the VDF measured on previous and subsequent peak-tracking scans.  The data processing algorithms flag full-scans so they can be easily identified by the data user, who may wish to remove or treat the full-scan spectra differently.  When deciding on the operating plan for SPC, the number of full-scans performed is kept to a minimum, but must be balanced with the advantages provided by full scans allowing SPC to accurately identify the peak of the distribution so that peak-tracking can be performed.  

%%%%%%%%%%%%%%%%%%%%%%%%%%%
\subsection{Typical On-orbit Operation}
Though the configuration parameters of SPC allow for a huge variety of different acquisition modes, the parameters are typically operationally restricted to one of the few configurations shown in Table \ref{tab:data_modes}.  A major reason for doing this is so that each spectrum begins precisely at the same time as each spectrum measured by the SPAN instruments and is synchronized to the magnetic and electric field measurements \citep{bale_2016}.

Generally, each PSP orbit is separated into two distinct regions: ``cruise'' (outside of 0.25 AU) and ``encounter'' (inside 0.25 AU) \citep{korreck_2014}.  During the cruise phase, the instruments are only operated sporadically, and at a fairly low data-rate (and thus a fairly low measurement cadence).  Typically, this is due to restrictions on power and data-rate that require the instruments to be turned off when the spacecraft is transmitting using its Ka-band high-gain antenna, or when a high-speed data transfer is occurring between an instrument and the spacecraft.  During the encounter phase, the instruments are powered continuously and data are acquired at a much higher cadence.  For SPC, the typical data acquisition modes are shown in Table \ref{tab:data_modes}.  The precise amount of time spent in each mode is different in each orbit and depends on the volume of data that is allocated to SPC and the expected signal-to-noise at the heliocentric distances for that particular orbit.

\startlongtable
\begin{deluxetable*}{c c c c c}
\tablecaption{Typical SPC Data Acquisition Modes \label{tab:data_modes}}
\tablehead{
\colhead{Mode Name} & \colhead{When Used} & \colhead{\# Voltage Windows} & \colhead{\# Spectra} & \colhead{Portion of} \\
\colhead{} & \colhead{} & \colhead{Per Spectrum} & \colhead{Per Packet\tablenotemark{a}} & \colhead{Packets Sent}
}
\startdata
Ion SuperFast   & Encounter (Extremely Rarely)	& 8     & 16    & all \\
Ion Fast        & Encounter         	    	& 30    & 4     & all \\
Ion Med-Fast    & Encounter              	    & 30    & 2     & all \\
Ion Medium      & Encounter              	    & 30    & 1     & all \\
Ion Slow        & Cruise (Typical)              & 30    & 1/4   & 4 out of 32 \\
Ion Flux-Angle  & Encounter (10 min per day)    & 1     & 64-256     & all \\
Electron Medium & Encounter (Rarely)        	& 12    & 1     & all \\
Electron Slow   & Cruise (Extremely Rarely)		& 30    & 1     & 1 out of 32
\enddata
\tablenotetext{a}{Packets are acquired over the course of 0.87381$\overline{3}$ seconds; this time period was selected due to the need to fit an integer power of 2 clock cycles into the measurement period (i.e., $2^{24}/19.2 MHz = 0.87381\overline{3}$ seconds.)}
\end{deluxetable*}

%%%%%%%%%%%%%%%%%%%%%%%%%%%%%%%%%%%%%%%%%%%%%%%%%%%%%%%
\section{Data Description} \label{sec:data_description}
The typical science packets from the SPC instrument contain the sine and cosine amplitudes from the best gain for each of the four collector plates, stored as 12-bit digital numbers representing a signed integer (i.e., a number between -1024 to +1023).  These data are processed further (onboard the SWEM) before being sent to the ground.  The only calculation performed by the SWEM is to take the root-sum-square (RSS) of the sine and cosine values: $RSS=1.414\sqrt{sine^2+cosine^2}$, which results in an unsigned 12-bit integer representing the AC magnitude of the current on each collector plate, while information about the phase of the signal is lost.  The advantage of this is a reduction in telemetry volume by approximately a factor of 2, with the negative effect of the addition of noise that is approximately $\sqrt{2}$ higher.  

The data from SPC are received as binary ``level-0'' files that have been downloaded from the spacecraft via the Deep-Space Network (DSN).  These files contain packets of data of different types (e.g., analog housekeeping, digital housekeeping, science.)  The packets are decoded and time-ordered to produce ``level-1'' files that are in the ``Common Data Format''\footnote{\url{https://cdf.gsfc.nasa.gov/}} (CDF)  \citep{treinish_1987}.  All of the level-1 science products are stored in ``digital number'' units, and as simple time series.  Because the level-1 files are not in physically meaningful units, these files are not distributed to the public.  The housekeeping data are converted into physically-meaningful units and stored in ``level-2'' files that are used to trend the health of the instrument.  The science data are processed into level-2 files (containing current flux spectra) and level-3 files (containing the results of moments and fits to the level-2 spectra: density, velocity, and thermal speed.)  The development of those level-2 and level-3 data products are described in the remainder of this section.

\subsection{Calibration of Data}
The level-1 files contain values that represent the measured peak-to-peak current in each collector plate, but include some contribution from noise sources that can be removed during data processing (e.g., signals from another collector plate inadvertently being measured).  For each of the four gain stages, there is a conversion from the digital number to a current (units of pico-amps) that requires knowledge of the relative response of each gain stage to the others as well as the absolute response of one of the gain stages.  Additionally, the absolute collection efficiency of the instrument must be known to convert the current to a current flux (units of pA/cm$^2$.)

The effective collection area, $A_{eff}$ of the instrument is the starting point for determining the absolute efficiency.  For a normally-incident parallel beam of particles (i.e., a particle population that is flowing perpendicular to the plane of the instrument aperture and with zero thermal velocity in the plane of the instrument aperture) this is a trivial calculation: $A_{eff}=A_{LA}T^{n}$, where $A_{LA}$ is the limiting aperture area, $T$ is the transparency of each single grid ($\sim$0.9) and $n$ is the number of grids (8).  For this simple case, $A_{eff}$ is 1.59 $cm^2$.  

The situation is more complex for off-axis flows and for particle populations with non-zero temperatures, in which case the detection efficiency must be determined as a function of the incidence angle of the ion flow.  This efficiency is determined through the use of a Monte-Carlo simulation that propagates test particles through the instrument geometry accounting for the transverse velocity of each particle, grid transparency as a function of angle, refraction of particles as they are slowed and sped up by the electric field near the modulator grid, gaps between the collector plates, and occultation of the incoming particle beam by the entrance aperture (for incidence angles outside of the nominal field-of-view.)  The total efficiency of the instrument is then determined by combining the effective collection area with the off-axis efficiency and scaling the result so that the SPC density measurements match those made by the independent measurement of electron density made by the FIELDS instrument (see Appendix \ref{sec:appendix:uncertainties} for details.)

\subsection{Converting to Level-2 Files}
Level-2 science products are created by applying the algorithms discussed above to convert the digital numbers into physically meaningful units.  Additionally, the measurements are sorted into spectra (i.e., each scan through a range of voltages is grouped together.)  The data are stored in CDF files and internally documented, including a description of each variable, its units, and more.  The SPC L2 CDF contains the following variables:
\begin{itemize}
    \item Epoch: the beginning time of each spectrum
    \item Measurement time: the time of the beginning of the measurement for each step in the spectrum
    \item Voltage (2x): the lower and upper bounds of the voltage window that was used for each step in the spectrum
    \item Current (4x): the current measured on each collector plate
    \item Flow-angle (2x): the angles at which the beam of particles enters the instrument
    \item Differential Energy Flux: the total current on all 4 collector plates combined and divided by the effective area of the instrument.  The differential energy flux can be converted into a 1-D distribution function in the velocity component normal to the sensor, $F(v_z)$, (a ``Reduced Distribution Function'', RDF), as described in Appendix \ref{sec:appendix:vdf}.
    \item Uncertainties (many): each variable listed above also has an associated uncertainty variable describing relative precision to which a given measurement is to be believed.
\end{itemize}

\begin{figure}[ht!]
\plotone{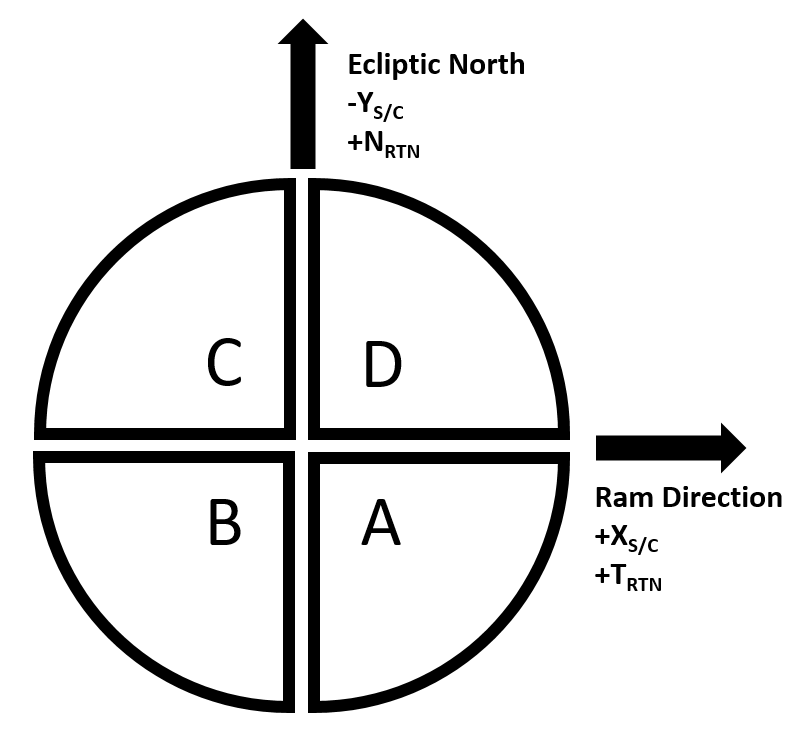}
\caption{Collector plate orientation, as viewed from the back of the instrument (i.e., looking toward the Sun.)   \label{fig:cp_orientation}}
\end{figure}

The flow angle shown in the L2 file is a measurement of the incidence angle of the solar wind beam for the E/q window with the highest current.  The angle is calculated by measuring the difference in current on pairs of collector plates (see Figure \ref{fig:cp_orientation} for the orientation of the collector plates).  For the currents on each of the four collector plates, $I_A, I_B, I_C, I_D$, the two flow angles are calculated using a linear approximation and assuming a cold plasma via the following equations:

\begin{eqnarray}
I_{TOT} = I_A + I_B + I_C + I_D\label{eqn:itot} \\
\phi = \lambda \frac{I_A + I_D - \left(I_B + I_C\right)}{I_{TOT}}\label{eqn:phi} \\
\theta = \lambda \frac{I_C + I_D - \left(I_A + I_B\right)}{I_{TOT}}\label{eqn:theta}
\end{eqnarray}

where, in spacecraft coordinates, $\phi$ is the flow angle in the X-Z plane and $\theta$ is the flow angle in the Y-Z plane, in radians. The constant $\lambda = (\pi/2) * (r_{LA}/d_{LA}) \approx 1.009$ is determined by the radius of the limiting aperture, $r_{LA}$, and its axial distance from the collector plates, $d_{LA}$ (see table \ref{tab:spc_sizes_locs}). The cold plasma approximation is not necessarily a good one for plasma distributions that have a particularly low Mach number.  For inflow mach numbers greater than 10, the linear approximation is good to within about 1-10\% within SPC's nominal field-of-view (30 degrees half-angle). The angular sensitivity of the instrument changes substantially from Mach 10 down to Mach 3, such that the linear approximation is only $\sim50\%$ accurate at Mach 3. Higher level data products are derived with a temperature-dependent model of the instrument's angular response in order to properly account for finite Mach number effects. \edit1{The model is a Monte-Carlo simulation of individual particle trajectories through the SPC cup geometry that is performed for a wide range of Mach numbers and incidence angles.}

\subsection{Converting to Level-3 Data}
Level-3 products are also publicly available.  Every ion spectrum in the level-2 file has a corresponding entry in the level-3 file.  The level-3 product contains the vector velocity, density, and [radial] thermal speed in two separate coordinate systems for multiple ion populations (one proton population, one alpha population, and one additional population of arbitrary charge/mass.)  The proton parameters are obtained in two separate ways, and the estimates derived from both methods are provided in the L3 files: firstly, moments of the reduced distribution function are taken over a limited range of energy/charge, and secondly, one or more Maxwellian distributions are fit to current spectra..

For the calculation of proton velocity moments, first the differential energy flux spectrum from the level-2 file is converted into a reduced distribution function (RDF) as per the method described in Appendix \ref{sec:appendix:vdf}.  Then, an appropriate energy range must be identified that contains only the solar wind protons. That range is generally found by identifying the locus of current measurements within each spectrum that exceeds a threshold signal-to-noise ratio (S/N) and includes the peak measurement.  Additionally, the energy range is narrowed to mask out energies where alpha particles would likely contribute significantly to the total current measured.

To compute velocity moments over such a range allows for the possibility of systematic errors because (1) proton and alpha particle measurements may overlap and (2) the tails of the distribution are signal-to-noise limited.  These errors can result in a too-narrow energy range that under-counts protons.  This is an unavoidable complication that arises in all solar wind ion detectors with no mass identification, and one for which a more accurate measurement of the proton parameters can only be derived from a VDF model fit of overlapping distributions rather than moments.  L3 data quality flags are included, where applicable, to indicate when the proton sub-range runs continuously into the masked out region.

Level-3 ion data products also include Maxwellian model fits to the level-2 current spectra. These fits are obtained by nonlinear least-squares regression of the measurements to an analytic model of the SPC instrument response to one or more isotropic Maxwellian distributions of inflowing ions. The populations typically include the primary proton peak, the primary alpha particle peak, and a third population that may be assigned an arbitrary ionic mass and charge, but is most often used to describe the secondary proton beam/shoulder.

An array of data quality flags with descriptive metadata accompanies each measurement in the level-3 products. Each array contains information regarding the success of the peak-fitting and any constraints that were applied to obtain convergence, as well as information regarding the degree of confusion between species, the completeness of the spectrum energy range, any indicators of reduced measurement quality, and other parameters that are fully described in the data files.  Fitting uncertainties are also provided with all parameters that are derived in this way.

\subsection{Temperature Anisotropy}
The solar wind often exhibits an anisotropic temperature, in which the temperature of the plasma in the direction perpendicular to the magnetic field ($T_{\perp}$) is higher or lower than the temperature in the direction parallel to the magnetic field ($T_{||}$).  This temperature anisotropy, $\alpha = T_{\perp}/T_{||}$, is useful for helping to understand the kinetic effects shaping the velocity distribution in the solar wind.  The SPAN-ion instrument is able to more easily measure temperature anisotropy, but for times when the solar wind flow is directed into the SPC aperture, it is useful to have a method for calculating $\alpha$ from SPC measurements alone.  

\deleted{With a Sun-pointing instrument such as SPC, this can be done in a couple of ways: 1) by waiting for the magnetic field to rotate sufficiently to provide multiple measurements of the radial distribution function along different angles through the 3-dimensional velocity distribution function \citep{kasper_2002}, or 2) by using the relative signal on each of the segments of the collector plates to calculate the anisotropy within a single measurement of the radial distribution function.  The latter method is described in the rest of this section with the hope that the instrument team will eventually be able to provide a temperature anisotropy data product for those time periods when SPC is the primary instrument measuring the solar wind.}

\added{With a Sun-pointing instrument such as SPC, calculating anisotropy is considerably more difficult than with instruments such as Wind/SWE \citep{ogilvie_1995}, where the spinning spacecraft provides the ability to regularly make measurements of the velocity distribution function at different angles with respect to the magnetic field \citep{kasper_2002}.  A similar analysis can be performed with SPC by accumulating measurements over a long enough time period that the magnetic field rotates sufficiently throughout SPC's field-of-view to provide measurements of the radial distribution function through different angular slices of the 3-dimensional velocity distribution function \citep[Huang et al., this issue][]{}.  This method requires that the plasma distribution does not change appreciably throughout the course of the measurement and inherently reduces the cadence with which the anisotropy can be measured.}

\added{A second method of determining anisotropy is} by using the relative signal on each of the segments of the collector plates to calculate the anisotropy within a single measurement of the radial distribution function. \added{The method makes use of the fact that the apparent flow angle of the plasma varies throughout a spectrum, depending on the anisotropy of the distribution.  This method fails when the magnetic field is nearly parallel or perpendicular to SPC's look direction, but allows for anisotropy to be calculated on the same cadence that spectra are being measured.} \replaced{The latter method is described in the rest of this section with the hope that}{Further development and testing of the algorithms used to calculate anisotropy based on these two methods will eventually allow} the instrument team \deleted{will eventually be able} to provide a temperature anisotropy data product for those time periods when SPC is the primary instrument measuring the solar wind.

\deleted{While SPC only samples the distribution function of the solar wind plasma along a single axis, each measurement along $v_z$ (perpendicular to the cup aperture) will be an integration over measurements in $v_x$ and $v_y$, which allows for temperature anisotropy to be calculated for periods when the magnetic field direction is not parallel or perpendicular to the cup axis.  The perpendicular components of the plasma flow direction at each $v_z$ are determined by the peaks of the distribution in the $(\hat{x},\hat{z})$ and $(\hat{y},\hat{z})$ planes. If the shape of the velocity distribution is ellipsoidal in either plane due to a temperature anisotropy, the instrument will measure a change in the flow direction across measurements in $v_z$. This change in the flow direction will be directly related to the anisotropy of the plasma temperature.}

\deleted{We assume a bi-Maxwellian particle distribution aligned with respect to the magnetic field and are thus able to infer the parallel and perpendicular plasma temperatures during each sampling of the VDF by SPC. By making use of the symmetry in the bi-Maxwellian distribution about B, we can show that for a particle distribution with anisotropy, $\alpha = T_{\perp}/T_{||}$, bulk velocity $\vec{u}$, and a VDF of $f(\vec{v})$:}

   \deleted{ \begin{equation}
        (v_x-u_x) = - \frac{(\alpha-1)\hat{B}_x}{\hat{B}_x^2(\alpha-1) +1} \left[ (v_y-u_y)\hat{B}_y + (v_z-u_z)\hat{B}_z \right]
    \end{equation}}
    
   \deleted{ \begin{equation}
        (v_y-u_y) = - \frac{(\alpha-1)\hat{B}_y}{\hat{B}_y^2(\alpha-1) +1} \left[ (v_x-u_x)\hat{B}_x + (v_z-u_z)\hat{B}_z \right]
    \end{equation}}

\deleted{which reduces to two independent relationships to $\alpha$}

    \deleted{\begin{equation}
    \label{eq:slopeDef}
        \{\chi,\psi\}
        =
        - \frac{(\alpha-1)\hat{B}_z}{(\alpha-1)(1-\hat{B}_z^2)+1}
    \end{equation}}
    
\deleted{where the slopes $\chi$ and $\psi$ given by}

    \deleted{\begin{equation}
        \label{eq:ChiPsi}
        \chi=\frac{(v_x-u_x)}{(v_z-u_z)\hat{B}_x}
        \hspace{0.5cm},\hspace{0.5cm}
        \psi=\frac{(v_y-u_y)}{(v_z-u_z)\hat{B}_y}
    \end{equation}}

\deleted{are determined from a linear fit to the flow direction measurement across the VDF peak. This method of determining the particle anisotropy is valid where the background magnetic field direction is not directed either purely parallel or perpendicular to the instrument look direction, i.e. $(B_z/|B|) \ne$  \{0,1\}. }

%%%%%%%%%%%%%%%%%%%%%%%%%%%%%%%%%%%%%%%%%%%%%%%%%%%%%%%
\section{Observed Performance} \label{sec:observed_performance}
The PSP mission was launched on 2018/08/12 and SPC was first powered up on 2018/08/30.  Directly following turn-on, SPC was put through a variety of activities known as ``instrument commissioning'' to assess its post-launch performance and prepare it for the first solar encounter.  These tests included calibration runs using its internal calibration source, measuring the high-voltage current draw on the high-voltage power supply, and a rotation of the spacecraft so that multiple instruments could alternately be pointed toward the solar wind.

\begin{figure*}[ht!]
\plotone{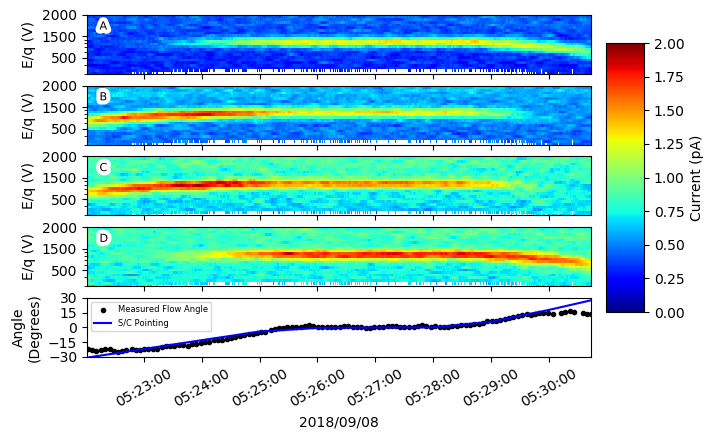}
\caption{Data taken during a spacecraft rotation during instrument commissioning.  As the spacecraft rotates, the beam of solar wind ions moves from collector plates B and C to collector plates A and D.  In the bottom panel, the calculated flow angle matches the spacecraft pointing angle for the time periods with high signal-to-noise. \label{fig:rotation}}
\end{figure*}

Figure \ref{fig:rotation} shows spectrograms from each of the four collector plates during a spacecraft rotation test during the commissioning period.  The instrument was acquiring one spectrum every 0.873 seconds.  To smooth the data and enhance the signal-to-noise ratio, each voltage window was averaged with the same voltage windows in the preceding and following five spectra.   At the beginning of the time period shown, the spacecraft was pointed -30 degrees off of the Sun-spacecraft line (as shown by the blue line in the bottom panel).  The spacecraft then slewed to a Sun-pointed attitude, where it stayed for about 3 minutes before continuing to slew away from the Sun in the opposite direction.

SPC is designed such that an incoming beam that is incident from off of the instrument symmetry axis produces a larger current in 2 of the collector plates.  In the case of the beginning of this test, collector plates B and C had a small, but measurable, signal and collector plates A and D were measuring only noise.  This measurement was made difficult due to the fact that signals were very small due to the spacecraft being 0.94 AU away from the Sun while the instrument was designed to work within 0.25 AU.  As the spacecraft rotated toward the Sun, the total signal increased (due to the \edit1{aperture being pointed more directly toward the solar wind source and to the} increasing transparency of the grids) and the signal was spread more evenly across all four collector plates.  The black points in the bottom panel of Figure \ref{fig:rotation} show the calculated flow angle from the four collector plate signals, as per Equations \ref{eqn:itot}-\ref{eqn:theta}.  The measured flow angle deviates from the spacecraft pointing angle when the signal gets too low, but in the higher signal-to-noise regions they match quite well.  This confirms the functionality and relative calibration of each collector plate.

\begin{figure*}[ht!]
\plotone{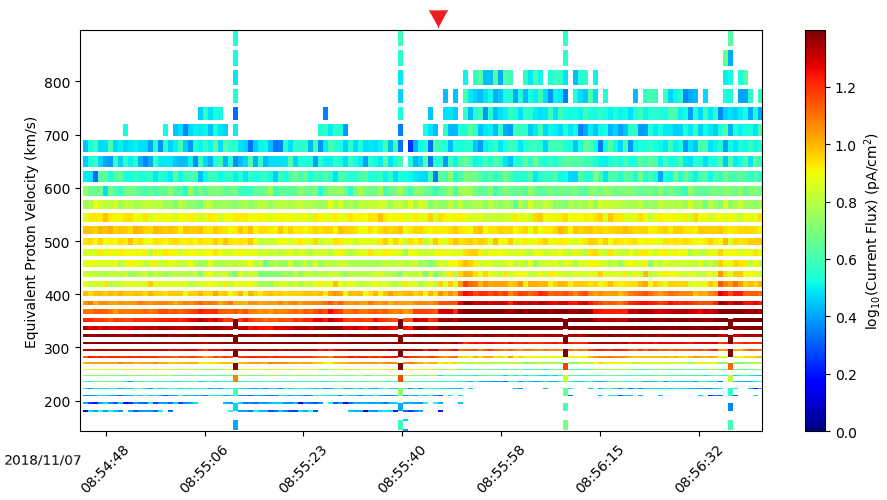}
\caption{Spectrogram measured during encounter \#1 for a period of time just before the first perihelion. The red triangle marks the spectrum that is shown in Figure \ref{fig:vdf_fit}. \label{fig:spectrogram}}
\end{figure*}

Figure \ref{fig:spectrogram} shows a representative spectrogram of the measured current flux that covers a period of time during the first encounter \edit1{when PSP was at 0.17 AU}.  Color shows the total current summed over all four collector plates and divided by the effective area, as a function of the equivalent proton velocity (v$^*$) and time, where $v^* =\sqrt{2V/(q_em_p)}$, \deleted{where}V is the modulator voltage, $q_e$ is the fundamental charge, and $m_p$ is the mass of a proton.  The primary proton peak is seen as the \added{horizontal} band of red pixels between 300 and 350 km \replaced{/s}{s$^{-1}$}.  \edit1{A second distinct peak can be seen as a horizontal band of orange pixels at approximately 525 km s$^{-1}$.  The secondary peak is most likely alpha particles that, in the units shown, would show up at $\sqrt{2}$ times the proton velocity if they were co-moving with the protons.  Thus, there is likely some differential flow between the alphas and protons during much of time period shown.  }

Full-scans can be seen as spectra that cover the full velocity range (138 to 875 km/s).  At the lower velocities (below about 375 km/s), the instrument used wider windows so that the energy resolution would be worse, but the full voltage range could be covered in a relatively short period of time.  Because of those wider windows, the current measured during those windows is approximately double that measured in the adjacent peak-tracking scans.  The peak-tracking scans can be seen in Figure \ref{fig:spectrogram} as spectra that cover a narrower range of velocity.  The center of the peak tracking scans can been seen to move around in velocity space to track the highest measured current.

\begin{figure*}[ht!]
\plotone{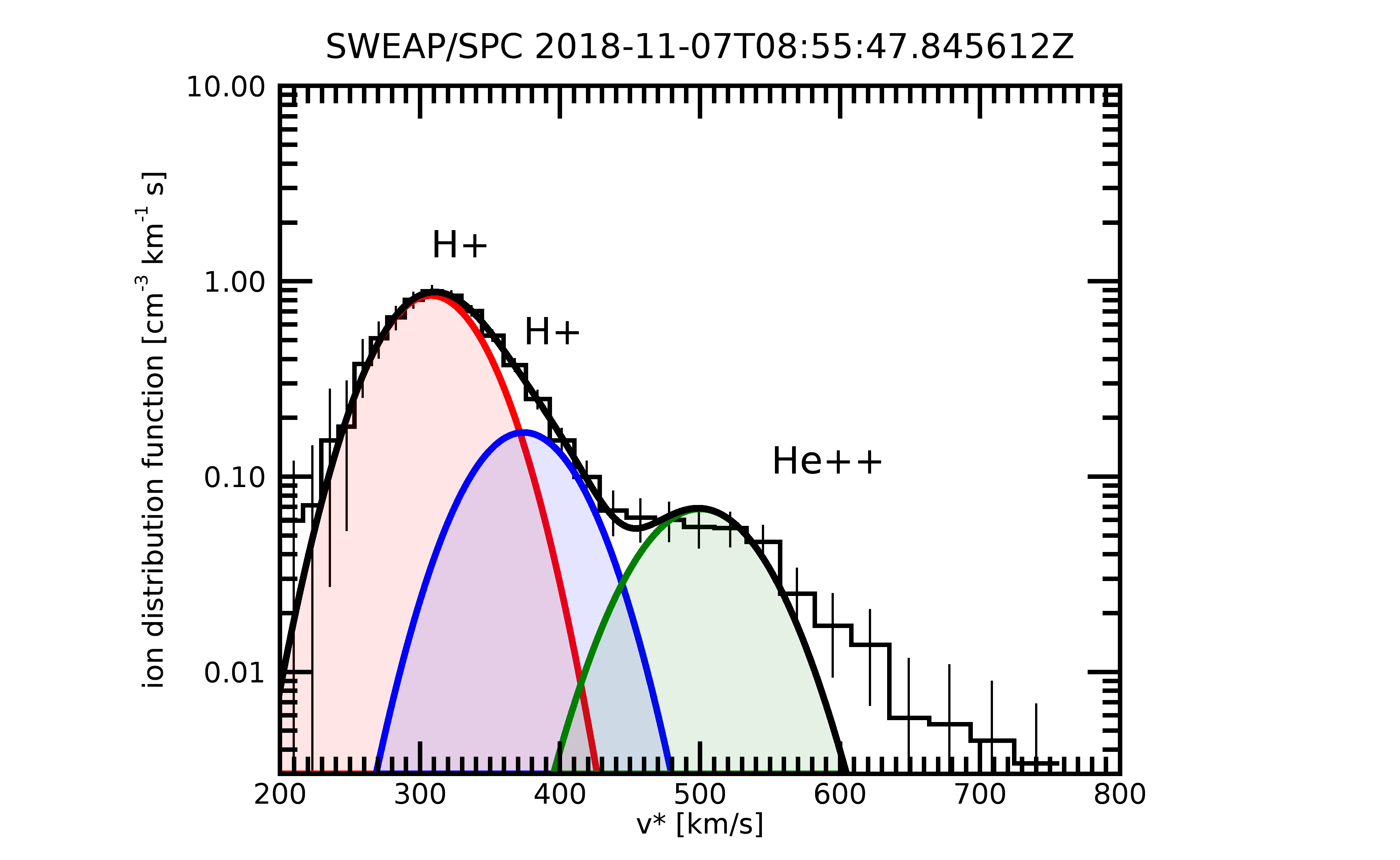}
\caption{Spectrum taken \edit1{on 2018/11/07 at 08:55:47,} during encounter \#1 just after perihelion.   The histogram line and error bars indicate the measured reduced distribution function and the uncertainties associated with each measurement.  The spectrum has been fit with Maxwellian distributions of core protons (red), a proton beam (blue), and an alpha particle population (green).  The total modeled distribution is shown as the smooth solid black line. \label{fig:vdf_fit}}
\end{figure*}

Figure \ref{fig:vdf_fit} shows a 1-D spectrum from \edit1{the time indicated with a red triangle at the top of Figure \ref{fig:spectrogram}}.  The differential energy flux has been converted to a reduced distribution function, as described in Appendix \ref{sec:appendix:vdf}.  The distribution function has been fit with three separate ion populations: a core proton population, a proton beam, and an alpha (He$^{++}$) population.  It was assumed that each population was a Maxwellian.  The signal-to-noise ratio in this particular spectrum is representative of typical spectra seen during encounter.  The alpha population in this spectrum is more clearly separable from the proton peak than is typical, which is due in this case to the relatively low temperatures of each population and to the relatively small proton-proton drift component normal to the sensor.

The SPC instrument was operated throughout the first two solar encounters of the Parker Solar Probe mission (here we define ``encounter'' to mean a continuous period of time in which the spacecraft is within 0.25 AU).  The first encounter occurred from 2018/10/31 to 2018/11/11, with its closest approach on 2018/11/06.  The second encounter occurred from 2019/03/30 to 2019/04/10, with its closest approach on 2019/04/04.  During the entirety of each of those encounters, the spacecraft was Sun-pointed, and the instrument was taking science measurements at a minimum of 1 spectrum per 0.873 seconds.  Outside of the encounter time periods, SPC was sporadically powered on to take science data at approximately 1 spectrum every 28 seconds.  Initial science results can be found in \citet{kasper_2019}.

Throughout the first encounter, the SPC instrument operated as expected, with high signal-to-noise ratios at closest approach (as in Figures \ref{fig:spectrogram} and \ref{fig:vdf_fit}.)  Data were also acquired successfully in electron mode, flux-angle mode, and at a range of different cadences to determine the best way to run the instrument for the remainder of the orbits.

\begin{figure*}[ht!]
\plotone{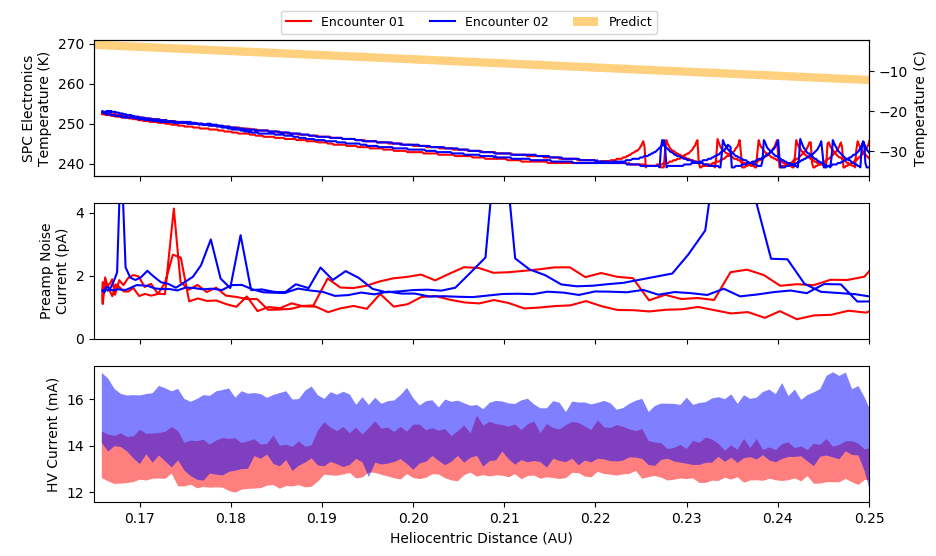}
\caption{Radial trends showing instrument performance over the first two encounters.  Red lines show data from encounter 1 and blue lines show data from encounter 2.  Top panel shows the temperature of the electronics box along with the range of pre-launch predictions of the electronics temperature (yellow shading).  The middle panel shows the pre-amplifier noise and the bottom panel shows the current draw on the high-voltage power supply.   \label{fig:radial}}
\end{figure*}

Figure \ref{fig:radial} shows radial trends of a few quantities that demonstrate the well-being of the instrument as it gets close to the Sun.  For each panel, the data from the first encounter is shown in red and the second encounter is shown in  blue.  The top panel shows the temperature of the electronics box along with a pre-flight prediction of the electronics box temperature (shown as a yellow band).  The electronics box temperature is the most important temperature in the instrument (and the only one measured) due to the maximum operating temperature of the electronic components (the instrument was qualified to operate up to 55 degrees Celsius.)  Outside of approximately 0.22 AU a thermostatically-controlled heater cycled on and off to keep the electronics box above its minimum operating temperature.  Inside of 0.22 AU, the temperature slowly increased as the spacecraft got closer to the Sun.  Generally, the observed electronics box temperature is 15-20 degrees below the pre-flight predictions.

The second panel of Figure \ref{fig:radial} shows a quantity designed to indicate the noise that is present on the front-end preamplifer circuit.  It is calculated from the level-2 spectra by averaging the five lowest values in the differential energy density spectrum in each full-scan over two hour span.  Both encounters show a minimal, if any, increase in noise as the spacecraft approaches the Sun.  The solar wind signal increases with decreasing heliocentric distance due to the approximately $1/r^2$ scaling of density with distance, resulting in a signal-to-noise ratio that increases as PSP gets closer to the Sun.  

The third panel of Figure \ref{fig:radial} shows the distribution of currents drawn from the high-voltage power supply at each radial distance.  The current during encounter 2 was noticeably higher than during encounter 1, most likely due to the fact that the maximum voltage used during encounter was 6 kV compared to 4 kV in encounter 1.  A pre-flight concern was that as the high-voltage insulators heat up when approaching the  Sun, they would become more conductive and lead to a higher current draw from the HVPS.  If the current draw were to get as high as about 100 mA, the instrument may be required to lower its maximum voltage.  As seen in Figure \ref{fig:radial} there is no apparent radial trend, and the current draw is at least a factor of five lower than the maximum allowed value.

%%%%%%%%%%%%%%%%%%%%%%%%%%%%%%%%%%%%%%%%%%%%%%%%%%%%%%%
\section{Concluding Comments}
A Sun-viewing solar wind ion instrument is essential for the fulfillment of all PSP science objectives, but major technological challenges had to be overcome in order to develop and test the instrument.  The need for SPC is even clearer after the first few encounters with the Sun and the discovery of large rotational flow of the solar wind that cancels out much of the abberation that was expected to allow the wind to flow into SPAN-Ai \citep{kasper_2019}.  Without SPC, observations of the core of the ion velocity distribution function would be limited to a handful of periods near the Sun.

High temperatures exceeding 1600C melt or evaporate typical materials, rule out standard manufacturing techniques, produce large thermal stresses, and break down most high voltage electrical insulators.  Many concerns were raised about the plasma near the Sun and the potential for sputtering, intense photo-electrons, runaway secondary electron emission, and other extreme versions of the space environment that have never been encountered before.  Specialized facilities had to be developed to test the instrument, and ultimately no single laboratory setup could reproduce the entire environment expected by PSP and SPC.  Ultimately the instrument passed these tests and has now shown that the high voltage insulator leakage current and the measurement noise are, if anything, weak functions of heliocentric distance.  The Solar Probe Cup is now poised to make high signal-to-noise observations of the solar wind and coronal plasma up through the final perihelia.

%%%%%%%%%%%%%%%%%%%%%%%%%%%%%%%%%%%%%%%%%%%%%%%%%%%%%%%
\appendix
\section{Deriving the Velocity Distribution Function} \label{sec:appendix:vdf}
The SPC level-2 data provide a variable called the ``Differential Charge Flux Density.'' In the following, we denote the particular ion species with mass $m$ and charge $q$, we denote the upper and lower voltages associated with a particular measurement as $V_{hi}$ and $V_{lo}$, respectively, and we denote a basis vector aligned with the SPC axis, into the cup, as $\hat{z}$. The Differential Charge Flux Density, which we denote $D$ gives the charge per second (in picoAmperes) that would cross a  $\hat{z}$-oriented 1 cm$^2$ unit area due to ions having kinetic energy in the range $V_{lo} \leq mv_z^2/2q \leq V_{hi}$.

If the ion species is known, it is straightforward to convert this value to a velocity distribution function through the process defined in this section.  In the case of SPC, where the velocity distribution function is only measured in one dimension (along the flow direction into the aperture of the instrument).  This is commonly called the ``reduced distribution function'', $F(v_z)$, and it can be approximately calculated with Equation \ref{eqn:rdf}. 

\begin{equation}\label{eqn:rdf}
    F(\overline{v}_z) \approx \frac{D}{q\overline{v}_z\Delta v},
\end{equation}

\noindent $\overline{v}_z$ is the average equivalent velocity of an ion during a voltage window measurement, and $\Delta v$ is the effective width of a voltage window in velocity units. The approximation becomes precise in the limit of small modulation amplitudes, i.e. $F'(v)\Delta v \ll F(v)$. Because of the sinusoidal nature of the modulating voltage, $\overline{v}_z$ must be calculated with Equation \ref{eqn:averagev}:

\begin{equation}\label{eqn:averagev}
    \overline{v}_z = \frac{2}{\pi}\sqrt{\frac{2qV_{hi}}{m}}E\left(\sqrt{\frac{V_{hi}-V_{lo}}{V_{hi}}}\right),
\end{equation}

\noindent where $V_{lo}$ and $V_{hi}$ are the lower and upper bounds of the voltage window, respectively, and $E()$ is the complete elliptic integral of the second kind.

The equivalent width of the voltage window in velocity space, $\Delta v$ is given by 

\begin{equation}\label{eqn:Deltav}
    \Delta v = \sqrt{ \frac{2q}{m}\left(V_{hi}+V_{lo}\right) -  \overline{v}_z^2}
\end{equation}

\section{General Calculations of Uncertainty in SPC Data Products} \label{sec:appendix:uncertainties}

Each SPC data file contains an uncertainty variable for each data variable in the file.  The uncertainties are calculated for each measurement, and should be used in any study making use of SPC data.  This appendix is not meant to provide uncertainties to be used in scientific calculations, but rather to give a general idea of the uncertainties you might expect from the SPC measurements.

SPC measures the flux of positive or negative charges, generally measured in picoAmperes, for charge carriers crossing a modulated voltage potential barrier. Uncertainties therefore follow from the electronics noise on the flux measurement and the fidelity of the voltage function, which typically amounted to 1-5$\%$ during the first two encounters. Series of flux as a function of voltage are converted to radial phase space distribution functions, which are integrated and/or fit to a peak model in order to estimate the densities, $n$, radial speeds, $v_R$, and temperatures, $T$, of the significant ions in the solar wind. 

During the first two encounters one finds that the solar wind protons are suitably described by a Maxwellian distribution function (as would be an ideal gas) about 60$\%$ of the time. Under those conditions, the precision with which $n$, $v_R$, and $T$ has been measured is estimated from the standard error on the best fit Maxwellian parameters. At SPC's native temporal resolution of about 4.6 Hz, typical uncertainties are $\sigma_{n}/n \approx 0.09$, $\sigma_{vR}/v_R \approx 0.03$, and $\sigma_{T}/T \approx 0.19$. 

When the Maxwellian model is not valid, the parameters are estimated by direct moment integration over the energy range where the protons are assumed to dominate the signal. The range over which the solar wind He$^{++}$ may be significant, which is expected to be roughly twice the voltage range associated with the proton peak, is excluded. For the purpose of estimating the uncertainty, however, the full range is considered. Asymmetric uncertainties are thus estimated based on (1) propagated measurement errors and (2) the truncation of the energy range over which the moment is calculated. At SPC's native temporal resolution of about 4.6 Hz, the 95$\%$ confidence intervals are typically $n_{95}\approx[0.95n, 1.33n]$, $v_{R,95}\approx[0.99v_R, 1.03v_R]$, and $T_{95}\approx[0.98T - 2T]$.

The off-radial components of the solar wind velocity are measured by comparing the relative fluxes measured at the four different quadrants of the SPC sensor. Due to the spread of the solar wind beam within the sensor, the angular uncertainty is temperature-dependent. For flow angles within the nominal field of view, approximately 30$^o$ half angle, the flux measurement uncertainty corresponds to angular precision of less than 1 degree in the cold plasma limit. At the temperatures and speeds observed over the first two encounters, the angular precision was typically 1-3 degrees. The median uncertainty for non-radial components of the proton velocity was about 9 km/s.

Consecutive measurements are independent, so under sufficiently steady conditions averaging down may be performed in order to trade temporal resolution for added precision in the usual way.

The absolute accuracy of the SPC radial speed and temperature, which is measured as a thermal speed, follow from the accuracy with which the modulated voltages are known. As verified in ground testing, the absolute accuracy for $v_R$ is less than 0.01$\%$ over a measurable range of approximately 119 km/s to 1065 km/s. The absolute accuracy in temperature is similarly negligible over a measurable range of approximately 7.3 kK to 21.1 MK (i.e. thermal speeds of 11 km/s to 600 km/s). Speeds and temperatures at the extremes of these ranges are subject to systematic considerations, but no such measurements have been presented here.

The accuracy of the density measurement follows from the effective sensitive area of the sensor, which is most accurately determined on orbit. The FIELDS experiment makes an independent measurement of the electron density, $n_e$, by performing quasi-thermal noise spectroscopy to determine the local electron plasma frequency \citep{bale_2016}. Similar complementary instruments have been used to calibrate the Wind Solar Wind Experiment \citep{maksimovic_1998, kasper_2006}. On PSP, $n_e$ is determined to arbitrary accuracy with a precision of about 6$\%$ when $n_e$ is large enough to provide a measurable signal. Under those conditions, the SPC sensitive area calibration is such that the median value of $n_p/n_e$ for SPC and FIELDS measurements performed within 1 minute of one another is $n_p/n_e\approx 0.97$. The remaining 3$\%$ of positive charge required for neutrality in the plasma is attributed to minor ions, primarily He$^{++}$. During the encounter period where this calibration could be performed, the sample was primarily slow solar wind. The He$^{++}$/H$^{+}$ abundance of the slow solar wind has been thoroughly studied by the \emph{Wind} mission \citep{kasper_2007}, suggesting that He$^{++}$/H$^{+}$  abundances of $0.5 - 2 \%$ by number (1-4$\%$ by charge) are to be expected. Thus the absolute accuracy of the SPC density measurement is estimated at $\approx 1\%$ and is no greater than 3$\%$.

The absolute accuracy for off-radial flow components follows from the inter-calibration of the four independent SPC sensor quadrants, each of which makes measurements over a sensitive range that is subdivided into four independent gain stages. The relative responses of the four quadrants and gain stages are normalized and verified via spacecraft roll maneuvers about the SPC symmetry axis. For solar wind fluxes typical of the first two encounters, the uncertainty associated with this calibration corresponds to a typical absolute accuracy of $\approx$0.5 degrees. The worst-case systematic error, which corresponds to flows near the edge of the nominal field of view and flux measurements at the lower extreme of the particular gain range is $\approx$1 degree.

\acknowledgments
This work was funded through work on the NASA contract \#NNN06AA01C.  The authors wish to acknowledge the significant work of all of the engineering staff that worked on the spacecraft and SPC instrument, especially Dick Gates, Henry Bergner, Chris Scholz, Matt Reinhart, Andrew Peddie, Shaughn London, Marc Musser, Todd Schneider, Jason Vaughn, and the staff at Oakridge National Laboratory.  Figure \ref{fig:spc_profile} photo courtesy of Andrew Wang.

\listofchanges{}

\end{document}